\documentclass[a4paper]{article}
\usepackage[margin=1in]{geometry} 

\usepackage{amsmath}
\usepackage{amsthm}
\usepackage{amssymb}
\usepackage{cite}

\usepackage[utf8]{inputenc}
\usepackage{hyperref}
\hypersetup{
	unicode,
	pdfauthor={Author One, Author Two, Author Three},
	pdftitle={A simple article template},
	pdfsubject={A simple article template},
	pdfkeywords={article, template, simple},
	pdfproducer={LaTeX},
	pdfcreator={pdflatex}
}

\usepackage{graphicx, color}
\graphicspath{{.}}

\usepackage{lipsum}

\title{A search for the first generation charged vector-like leptons at future colliders}
\author{Ali Osman Acar$^1$, Osman Emre Delialioglu$^1$, Saleh Sultansoy$^{1,2}$}

\date{
	\small{$^1$TOBB University of Economics and Technology, Ankara, Turkey \\ 
	$^2$ANAS Institute of Physics, Baku, Azerbaijan \\
}
}

\begin{document}
	\maketitle
	
	\begin{abstract}
	Flavor Democracy Hypothesis favors existence of iso-singlet quarks and vector-like
charged leptons. Their observation at future colliders could shed light on the nature
of mass and mixing patterns of known leptons and quarks, as well as Higgs boson
itself. Vector-like quarks are extensively searched by ATLAS and CMS
collaborations. Unfortunately, this is not the case for vector-like leptons, while they
have actually similar status from phenomenology viewpoint. We argue that vector-like
charged leptons should be included into the new physics search programs of future
energy frontier colliders. It is shown that pair production at the HE-LHC with decay of
one of the leptons to $Ze$ and another to $He$ channel, followed by $H \rightarrow bb$ and $Z \rightarrow \mu^+ \mu^-$ decays
will give opportunity to scan masses of iso-singlet and iso-doublet charged leptons up to 0.9
TeV and 1.5 TeV, respectively. FCC will extend this region up to 2 TeV for iso-singlet and
3.6 TeV for iso-doublet charged leptons.
		
	\end{abstract}

\tableofcontents

	\section{Introduction}
	\label{sec:intro}
The verification of the basic elements of the Standard Model has been completed with the discovery of the Higgs boson in the LHC ATLAS and CMS experiments \cite{aad2012observation,chatrchyan2012observation}. Nevertheless, there are a lot of problems which are not solved by the SM (see e.g. reviews \cite{sultansoy1998four,ssultansoytalk1} and references therein).Among them two most important ones, in our opinion, are: mass and mixing patterns of the SM fermions \cite{mchalm} in electroweak part and confinement \cite{de2017archeology} in QCD part. An essential stage in solving the first problem may be provided by the Flavor Democracy Hypothesis (see review \cite{sultansoy2007flavor} and references therein). The Confinement Hypothesis, in our opinion, will be clarified by future energy-frontier lepton-hadron colliders \cite{cetin2013qcd,fernandez2012large,doi:10.1142/S0217751X10049165,Acar:2016rde,canbay2017sppc,sdufeffd468814,kaya2019main} rather than theoretically (lattice \cite{greensite2003confinement} etc.).\\\\ Flavor Democracy (FD) Hypothesis has a half-century history (for details see Section 2). Flavor Democracy in three SM family case was excluded by the large value of the top quark mass. In order to preserve FD in the SM framework existence of the fourth SM family (SM4) was proposed in 1990s. However, minimal SM4 with one Higgs doublet is excluded by experimental Higgs data. On the other hand, Flavor Democracy may be preserved with introduction of heavy vector-like quarks and leptons. Today vector-like quarks are extensively searched by ATLAS and CMS collaborations, but this is not the case for vector-like leptons (VLL).\\\\
In this paper potential of future energy-frontier colliders for search for first generation charged vector-like lepton has been estimated. In section 2 we present basics of Flavor Democracy Hypothesis and its relation to vector-like quarks and leptons. Decay modes and production cross-sections for iso-singlet and iso-doublet VLL’s are considered in section 3. Section 4 is devoted to search for pair produced charged VLL at future colliders, where $L_{e} \rightarrow eZ/eH$ modes are considered with subsequent $H \rightarrow bb$ and $Z \rightarrow \mu^+\mu^-$ decays. Our conclusions and recommendations are given in section 5.  

	\section{Flavor democracy calls for iso-singlet quarks and vector-like leptons}
	\subsection{Weinberg’s statement}
Mass and mixing patterns of the SM fermions are among the most important issues, which should be clarified in particle physics. In recent interview published in CERN Courier \cite{mchalm} Steven Weinberg emphasized this point: “Asked what single mystery, if he could choose, he would like to see solved in his lifetime, Weinberg doesn’t have to think for long: he wants to be able to explain the observed pattern of quark and lepton masses”. What Weinberg meant can be understood from Table 1, where current values of the SM charged leptons and quarks are presented. One can see that the top quark mass is of order of electroweak scale, whereas masses of remaining SM fermions are much smaller. In our opinion, Flavor Democracy (see reviews \cite{sultansoy1998four,ssultansoytalk1,sultansoy2007flavor,arik2003turkish,sultansoy2012fourth,sultansoy2019energy} and references therein) could provide an important key to solve this mystery. 

\renewcommand{\arraystretch}{1.5}
	\begin{table}[ht]
		\centering
		\caption{Mass pattern of charged leptons and quarks}
		\begin{tabular}{|c|c|c|c|}
			\hline
			\textbf {} & \textbf{charged leptons} & \textbf{Up type quarks} & \textbf{Down type quarks}  \\
			\hline \hline
			$1^{st}$ Family & 0.5109989461 $\pm$ 0.0000000031  MeV & $2.16^{+0.49}_{-0.26}$ MeV & $4.67^{+0.48}_{-0.17}$ MeV \\
			\hline
			$2^{nd}$ Family & 105.6583745 ± 0.0000024 MeV  & 1.27 $\pm$ 0.02 GeV & $93^{+11}_{-5}$ MeV \\
			\hline
			$3^{rd}$ Family & 1776.86 ± 0.12 MeV & 172.76 $\pm$ 0.30 GeV & $4.18^{+0.03}_{-0.02}$ GeV \\
			\hline
		\end{tabular}
		\label{tbl:1}
	\end{table}

	\subsection{Flavor Democracy}
Flavor Democracy Hypothesis in framework of three family SM (SM3) was suggested in 1978 \cite{HARARI1978459}. As mentioned in Introduction, Flavor Democracy in three SM family case was excluded by the large value of the top quark mass and in order to preserve FD in the SM framework existence of the fourth SM family (SM4) was proposed in 1990s \cite{FRITZSCH199292,datta1993flavour,CELIKEL1995257}. However, minimal SM4 with one Higgs doublet is excluded by experimental Higgs data. 
On the other hand, Flavor Democracy may be preserved with introduction of heavy vector-like quarks and leptons \cite{kaya2018mass}. In addition, FD has essential consequences if applied to MSSM and preonic models \cite{sultansoy2007flavor}.
The next subsections (2.3-2.5) are based on reference \cite{kaya2018mass}.

 \subsection{Status of the Chiral Fourth Family}
It is known that the Standard Model does not fix the number of fermion families. This number should be less than 9 in order to preserve asymptotic freedom and more than 2 in order to provide CP violation. According to the LEP data on Z decays, number of chiral families with light neutrinos ($m_\nu << m_Z$) is equal to 3, whereas extra families with heavy neutrinos are not forbidden. The fourth chiral family was widely discussed thirty years ago (see, for example \cite{cline1,cline2}). However, the topic was pushed off the agenda due to the misinterpretation of the LEP data.\\
\\
Twenty years later 3 workshops on the fourth SM family \cite{beyond1,beyond2,beyond3} were held (for summary of the first and third workshops see \cite{holdom2009four} and \cite{cetin2011status}, respectively). Main motivation was Flavor Democracy \cite{FRITZSCH199292,datta1993flavour,CELIKEL1995257}, which naturally provides heavy fourth family fermions including neutrino (consequences of Flavor Democracy Hypothesis for different models, including MSSM and E6 inspired extension of the SM, have been considered in \cite{sultansoy2007flavor,sultansoy2000four}). In addition, fourth family gives opportunity to explain baryon asymmetry of universe; it can accommodate emerging possible hints of new physics in rare decays of heavy mesons etc. (see \cite{holdom2009four} and references therein). Phenomenological papers on direct production (including anomalous resonant production) of the SM4 fermions at different colliders are reviewed in \cite{PhysRevD.83.054022} (see tables VI and VII in \cite{PhysRevD.83.054022}).\\\\
This activity has almost ended due to misinterpretation of the LHC data on the Higgs decays. It should be emphasized that these data exclude the minimal SM4 with one Higgs doublet, whereas non-minimal SM4 with extended Higgs sector is still allowed \cite{bar2013two,banerjee2014higgs}. On the other hand, partial wave unitarity puts an upper limit around 700 GeV on the masses of fourth SM family quarks \cite{CHANOWITZ1979402}, which are excluded by the ATLAS and CMS data on search for pair u4 production \cite{atlas2015search,201882}.\\\\
Even if non-minimal SM4 may be excluded by the LHC soon, this is not the case for the general chiral fourth family (C4F). Therefore, ATLAS and CMS should continue a search for C4F up to kinematical limits. Concerning pair production, rescaling of the ATLAS lower bound using collider reach framework \cite{salam1} shows that LHC will give opportunity to cover Mu4 up to 1.6 and 2.5 TeV with integrated luminosities 300 and 3000 fb$^{-1}$, respectively. Resonant production of u4 via possible anomalous interactions may extend covered mass region up to 6 TeV \cite{beser2016possible}.

	\subsection{Iso-singlet quarks}
As mentioned in \cite{sultansoy1998four,ssultansoytalk1,sultansoy2007flavor}, large difference between $m_t$ and $m_b$ ($m_t >> m_b$) can be explained by the existence of iso-singlet down quarks predicted by E6 GUT. Unlike this, the presence of vector-like up quarks predicted by Little Higgs models is useless, since it leads to $m_b >> m_t$. Here we consider an addition of one iso-singlet down quark, so the quark sector is determined as 
	\begin{equation}
	\binom{u_L}{d_L},\binom{c_L}{s_L},\binom{t_L}{b_L},u_R,d_R,c_R,s_R,t_R,b_R,D_L,D_R
	\end{equation}
where D denotes new iso-singlet quark.\\\\
In the case of full Flavor Democracy, the mass matrix of the up type quarks can be written as \\
\begin{equation}
\centering
\begin{tabular}{cccc}
 & $u_R$ & $c_R$ & $t_R$ \\
$u_L$ & $a\eta$ &  $a\eta$ &  $a\eta$  \\
$c_L$ & $a\eta$ &  $a\eta$ &  $a\eta$ \\
$t_L$ & $a\eta$ &   $a\eta$  &  $a\eta$\\
\end{tabular}
\end{equation}
and mass matrix of down type quarks is \\
\begin{equation}
\centering
\begin{tabular}{ccccc}
 & $d_R$ & $s_R$ & $b_R$ & $D_R$ \\
$d_L$ & $a\eta$ &  $a\eta$ &  $a\eta$  &  $a\eta$ \\
$s_L$ & $a\eta$ &  $a\eta$ &  $a\eta$ &  $a\eta$ \\
$b_L$ & $a\eta$ &   $a\eta$  &  $a\eta$ &  $a\eta$ \\
$D_L$ & $M$ &   $M$  &  $M$ &  $M$ \\
\end{tabular}
\end{equation}
where $M$ ($M >> \eta$) is the new physics scale that determines the mass of iso-singlet quark. Diagonalization of matrix 2 and 3 results in $m_u=m_c=0$  and $m_t=3a\eta$ for up type quarks, $m_d=m_b=m_s=0$ and $m_D=3a\eta+M=m_t+M$ for down type quarks.\\\\
In order to obtain mass of b quark, small deviation from matrix (3) is involved, namely
\begin{equation}
\centering
\begin{tabular}{ccccc}
 & $d_R$ & $s_R$ & $b_R$ & $D_R$ \\
$d_L$ & $a\eta$ &  $a\eta$ &  $a\eta$  &  $(1-\alpha_b)a\eta$ \\
$s_L$ & $a\eta$ &  $a\eta$ &  $a\eta$ &   $(1-\alpha_b)a\eta$ \\
$b_L$ & $a\eta$ &   $a\eta$  &  $a\eta$ &   $(1-\alpha_b)a\eta$ \\
$D_L$ & $(1-\beta_b)M$ &   $(1-\beta_b)M$  &  $(1-\beta_b)M$ &  $M$ \\
\end{tabular}
\end{equation}
At this stage, for numerical calculations we assume $\alpha_b=\beta_b << 1$. While d and s quarks remain massless, we obtain following expressions for $\alpha_b$ and $m_D$
	\begin{equation}
	\alpha_b=\frac{(m_t+M)m_b}{2m_tM}
	\end{equation}
	\begin{equation}
	m_D \approx M+m_t-m_b
	\end{equation}
i.e. $\alpha_b \approx 1.31*10^{-2}$ and $m_D \approx 2169$ GeV if $M = $ 2 TeV.
Because the masses of $u$ and $d$ quarks are very small, we do not comment on them at this stage. Masses of $s$ and $c$ quarks can also be obtained due to small deviations from full democracy. Concerning $c$ quark let us consider following modification of the mass matrix of up quarks 
\begin{equation}
\centering
\begin{tabular}{cccc}
 & $u_R$ & $c_R$ & $t_R$ \\
$u_L$ & $a\eta$ &  $a\eta$ &  $a\eta$  \\
$c_L$ & $a\eta$ &  $a\eta$ &  $a\eta$ \\
$t_L$ & $a\eta$ &   $a\eta$  &  $(1+\alpha_c)a\eta$\\
\end{tabular}
\end{equation}
While $u$ quark remains massless, we obtain following expressions for $\alpha_c$ 
	\begin{equation}
	\alpha_c=\frac{9m_c}{2m_t}=3.3*10^{-2}
	\end{equation}
In order to obtain s quark mass, we consider following modification of the Eq. 4
\begin{equation}
\centering
\begin{tabular}{ccccc}
 & $d_R$ & $s_R$ & $b_R$ & $D_R$ \\
$d_L$ & $a\eta$ &  $a\eta$ &  $a\eta$  &  $(1-\alpha_b)a\eta$ \\
$s_L$ & $a\eta$ &  $a\eta$ &  $a\eta$ &   $(1-\alpha_b)a\eta$ \\
$b_L$ & $a\eta$ &   $a\eta$  &  $(1+\alpha_s)a\eta$ &   $(1-\alpha_b)a\eta$ \\
$D_L$ & $(1-\beta_b)M$ &   $(1-\beta_b)M$  &  $(1-\beta_b)M$ &  $M$ \\
\end{tabular}
\end{equation}
For $M = $ 2000 GeV,  $\alpha_b=\beta_b=1.32*10^{-2}$ and $\alpha_s=2.48*10^{-4}$ we obtain
	\begin{equation}
	m_D=2168\ \text{GeV}, m_b=4.18\ \text{GeV}, m_s=95.2\ \text{MeV}
	\end{equation}
Let us mention that down type iso-singlet quark give opportunity to explain $4\sigma$ discrepancy in the first road of CKM matrix (see \cite{belfatto2021ckm} and references therein).\\\\
Search for E6 iso-singlet quarks at the LHC was proposed in \cite{mehdiyev2007prospects,sultansoy2008e6,mehdiyev2008down,aguilar2009identifying}. According to PDG \cite{10.1093/ptep/ptaa104} $m_D > 1130$ GeV if B($D \rightarrow Zb$) = 1 from CMS and $m_D > 13500$ GeV if B($D \rightarrow Wt$) = 1 from ATLAS data.

	 \subsection{Vector-like leptons}
	Similarly to $b$ quark mass, low value of $\tau$ lepton mass can be provided by adding an isosinglet charged lepton
	\begin{equation}
	\binom{e_L}{\nu_{e_L}},\binom{\mu_L}{\nu_{\mu_L}},\binom{\tau_L}{\nu_{\tau_L}},e_R,\nu_{e_R},\mu_R,\nu_{\mu_R},\tau_R,\nu_{\tau_R},L_L,L_R
	\end{equation}\\
	Then, the mass matrix becomes as below
\begin{equation}
\centering
\begin{tabular}{ccccc}
 & $e_R$ & $\mu_R$ & $\tau_R$ & $L_R$ \\
$e_L$ & $a\eta$ &  $a\eta$ &  $a\eta$  &  $(1-\alpha_\tau)a\eta$ \\
$\mu_L$ & $a\eta$ &  $a\eta$ &  $a\eta$ &   $(1-\alpha_\tau)a\eta$ \\
$\tau_L$ & $a\eta$ &   $a\eta$  &  $a\eta$ &   $(1-\alpha_\tau)a\eta$ \\
$L_L$ & $(1-\beta_\tau)M$ &   $(1-\beta_\tau)M$  &  $(1-\beta_\tau)M$ &  $M$ \\
\end{tabular}
\end{equation}
At this stage, for numerical calculations we assume $\alpha_\tau=\beta_\tau << 1$. While electron and muon remain massless, we obtain following expressions for $\alpha_\tau$ and $m_L$
	\begin{equation}
	\alpha_\tau=\frac{(m_t+M)m_\tau}{2m_t M}
	\end{equation}
	\begin{equation}
	m_L \approx M+m_t-m_\tau
	\end{equation}
i.e. $\alpha_\tau \approx 6.02*10^{-3}$ and $m_L \approx 1171$ GeV if $M = 1$ TeV.\\\\
In order to obtain muon mass, we consider following modification of the Eq. 12
	
\begin{equation}
\centering
\begin{tabular}{ccccc}
 & $e_R$ & $\mu_R$ & $\tau_R$ & $L_R$ \\
$e_L$ & $a\eta$ &  $a\eta$ &  $a\eta$  &  $(1-\alpha_\tau)a\eta$ \\
$\mu_L$ & $a\eta$ &  $a\eta$ &  $a\eta$ &   $(1-\alpha_\tau)a\eta$ \\
$\tau_L$ & $a\eta$ &   $a\eta$  &  $(1+\alpha_\mu)a\eta$ &   $(1-\alpha_\tau)a\eta$ \\
$L_L$ & $(1-\beta_\tau)M$ &   $(1-\beta_\tau)M$  &  $(1-\beta_\tau)M$ &  $M$ \\
\end{tabular}
\end{equation}\\
For $M= 2000$ GeV, $\alpha_\tau=\beta_\tau=5.58*10^{-3}$ and $\alpha_\mu=2.73*10^{-4}$ this mass matrix lead to
	\begin{equation}
	m_L=2171\ \text{GeV}, m_\tau=1.777\ \text{GeV}, m_\mu=104.7\ \text{MeV}
	\end{equation}
Concerning neutrinos, small value of their masses may be provided by see-saw mechanism. If neutrino masses have Dirac nature solution can be provided by addition of isosinglet heavy neutrino.\\\\ In fact, it seems more natural to add new particles for each family. For example, in the E6 model an iso-singlet quark and vector-like lepton iso-doublet are added for each SM family leading to 6×6 mass matrices for down quarks (see \cite{sultansoy1998four,ssultansoytalk1,sultansoy2007flavor}) and charged leptons. Therefore, it is quite possible that intra-family mixings between SM and VL leptons are dominant. So far we have considered minimal case, namely, one new down quark and one new charged lepton have been added.
	\section{Charged vector-like lepton decays and production}
According to PDG \cite{10.1093/ptep/ptaa104} $m_L > 100.8$ GeV if B($L\rightarrow W\nu$) = 1 from LEP data. Here we are concerned with iso-singlet leptons and E6 GUT predicted iso-doublet leptons (for most general case, namely, left-handed n-plet and right-handed m-plet see \cite{heavysaleh}). Below we assume that the mixing of the new lepton with the electron is dominant. The situation where the mixing with the muon is dominant will give similar results with the $e-\mu$ exchange. The case in which the mixing with tau-lepton is dominant was examined in \cite{bhattiprolu2019prospects}.
	\subsection{Charged vector-like lepton decays}
	In the iso-singlet case, there is one mixing angle $\phi$ and 3 decay channels, namely, $W\nu_e$, $Ze$ and $He$. Branching ratios to these channels are presented in Fig. 1 (left).
\begin{figure}[!h]
  \centering
  \begin{minipage}[b]{0.49\textwidth}
    \includegraphics[width=\textwidth]{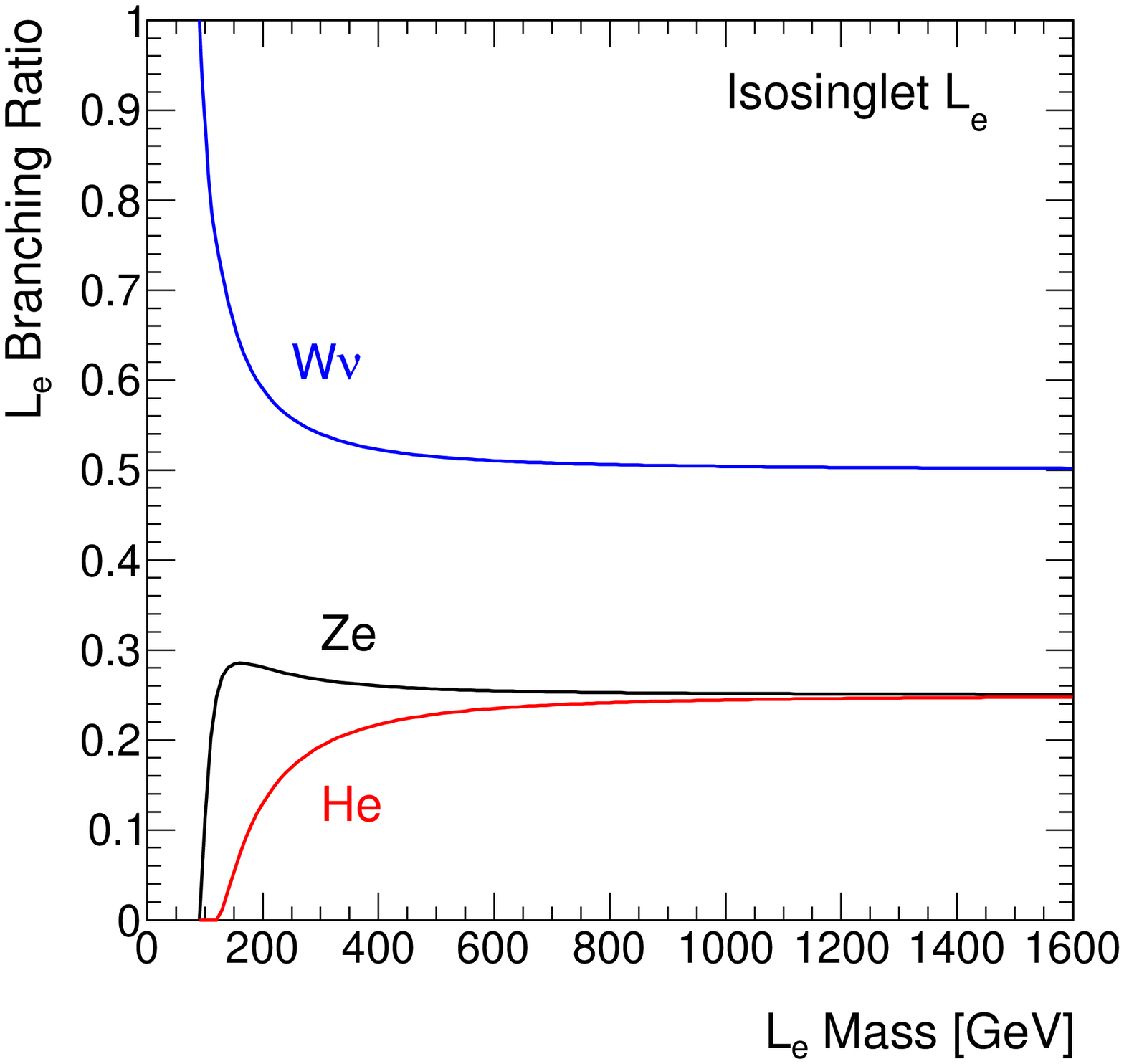}
  \end{minipage}
  \hfill
  \begin{minipage}[b]{0.49\textwidth}
    \includegraphics[width=\textwidth]{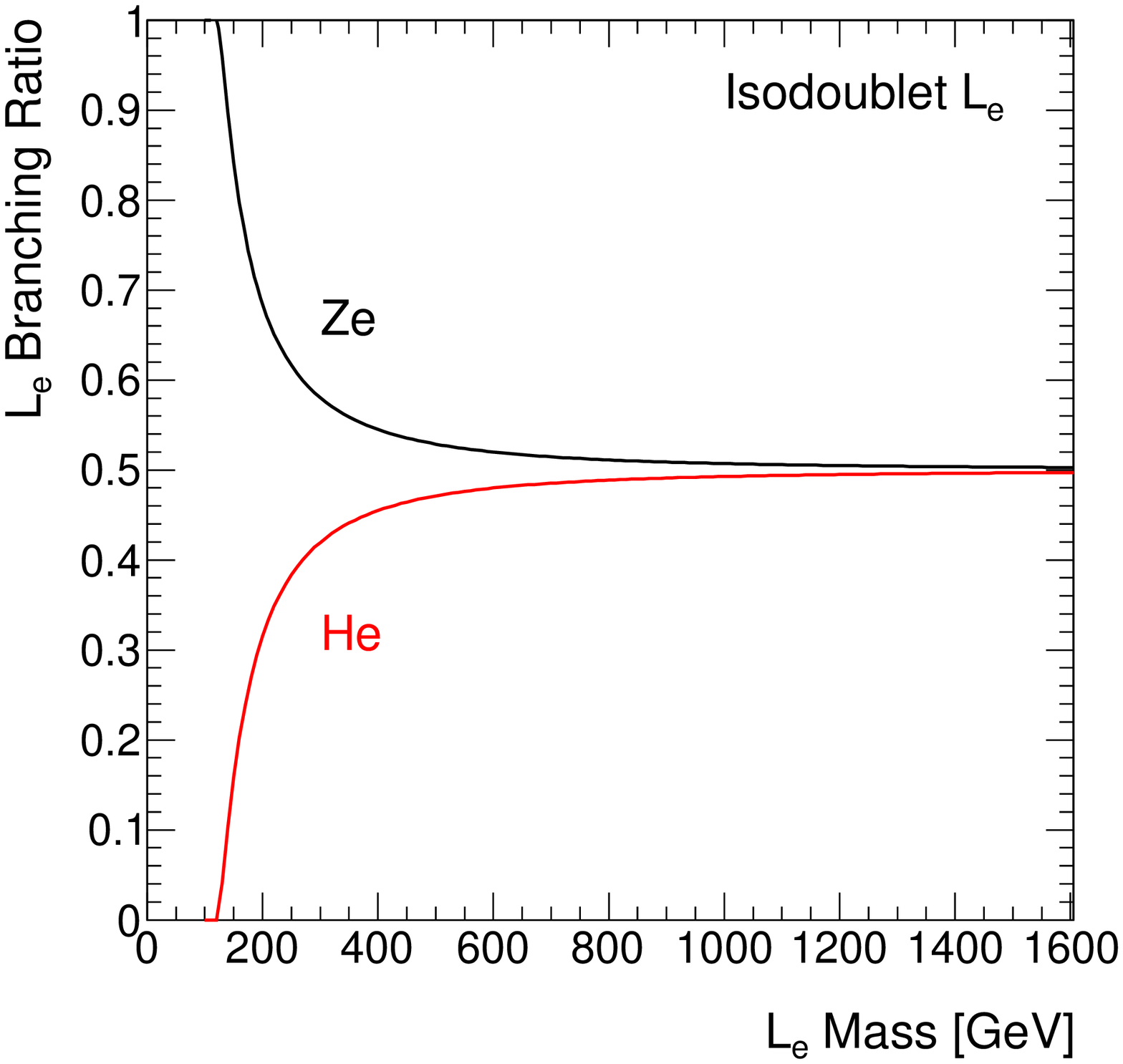}
  \end{minipage}
\caption{Branching Ratios for decays of iso-singlet (left) and iso-doublet (right, NC only) charged vector-like lepton.}
\end{figure}
 Electroweak precision data leads to $sin\phi < 0.018$ \cite{del2008effects}. In the iso-doublet case, we deal with two mixing angles: $\phi_R$, which is responsible for $Ze$ and $He$ decays, and $\phi_L$, which is responsible for $W\nu_e$ decay. In Fig. 1 (right) neutral current (NC only) branching ratios are shown assuming $sin\phi_L = 0$. If $sin\phi_L \approx sin\phi_R$ situation is similar to iso-singlet case presented in Fig. 1 (left).
	\subsection{Charged vector-like lepton production}
	Cross-sections for pair production of charged vector-like leptons at the LHC, HE-LHC and FCC are shown in Fig. 2 and Fig. 3, respectively. It is seen that higher mass values for iso-doublet leptons comparing to iso-singlet ones can be probed at hadron colliders. As for lepton colliders, the area that can be scanned for iso-singlet and iso-doublet masses is the same (and close to kinematical limit, $\sqrt{s}/2$).
\begin{figure}[!h]
  \centering
  \begin{minipage}[b]{0.49\textwidth}
    \includegraphics[width=\textwidth]{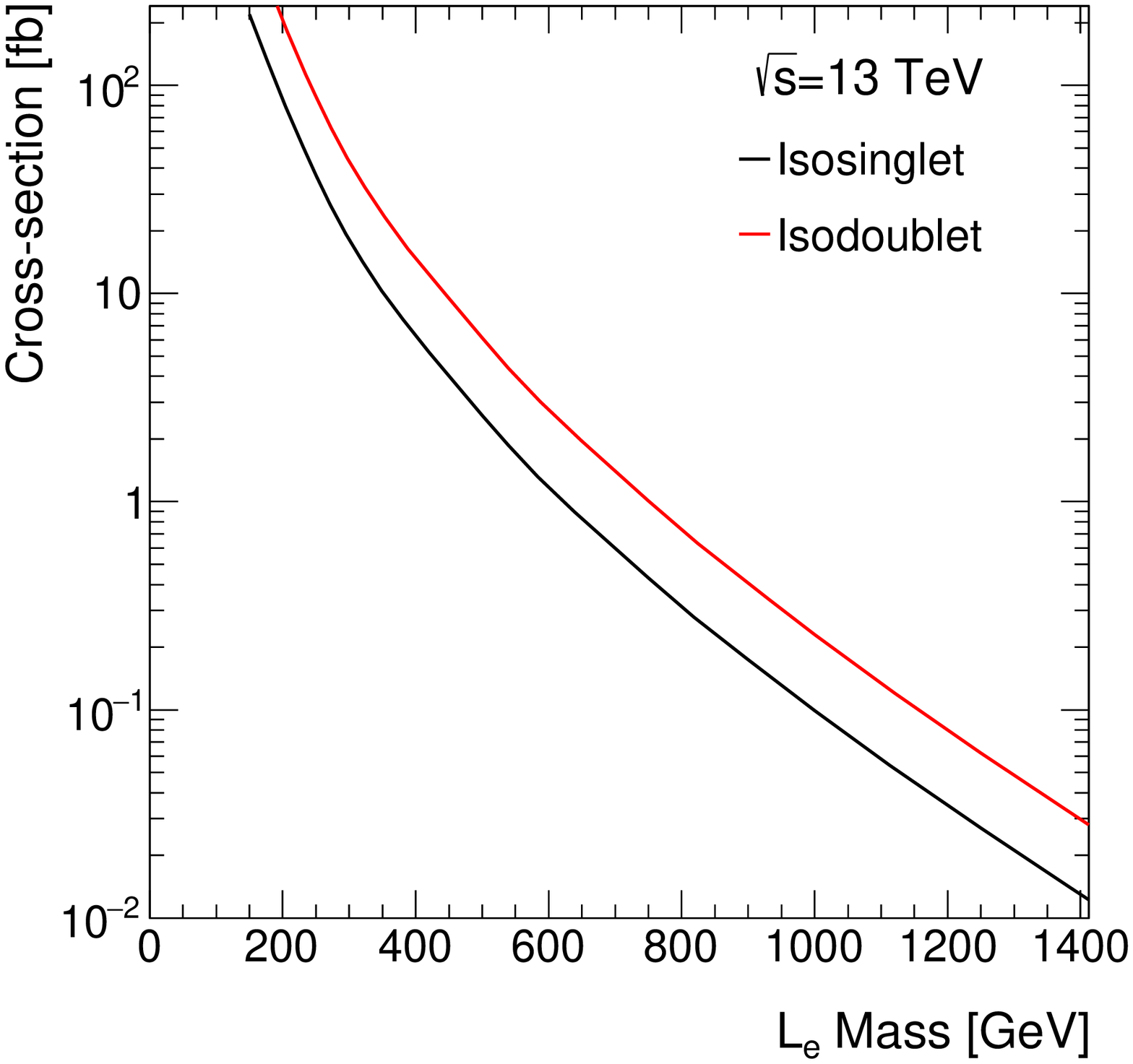}
  \end{minipage}
  \hfill
  \begin{minipage}[b]{0.49\textwidth}
    \includegraphics[width=\textwidth]{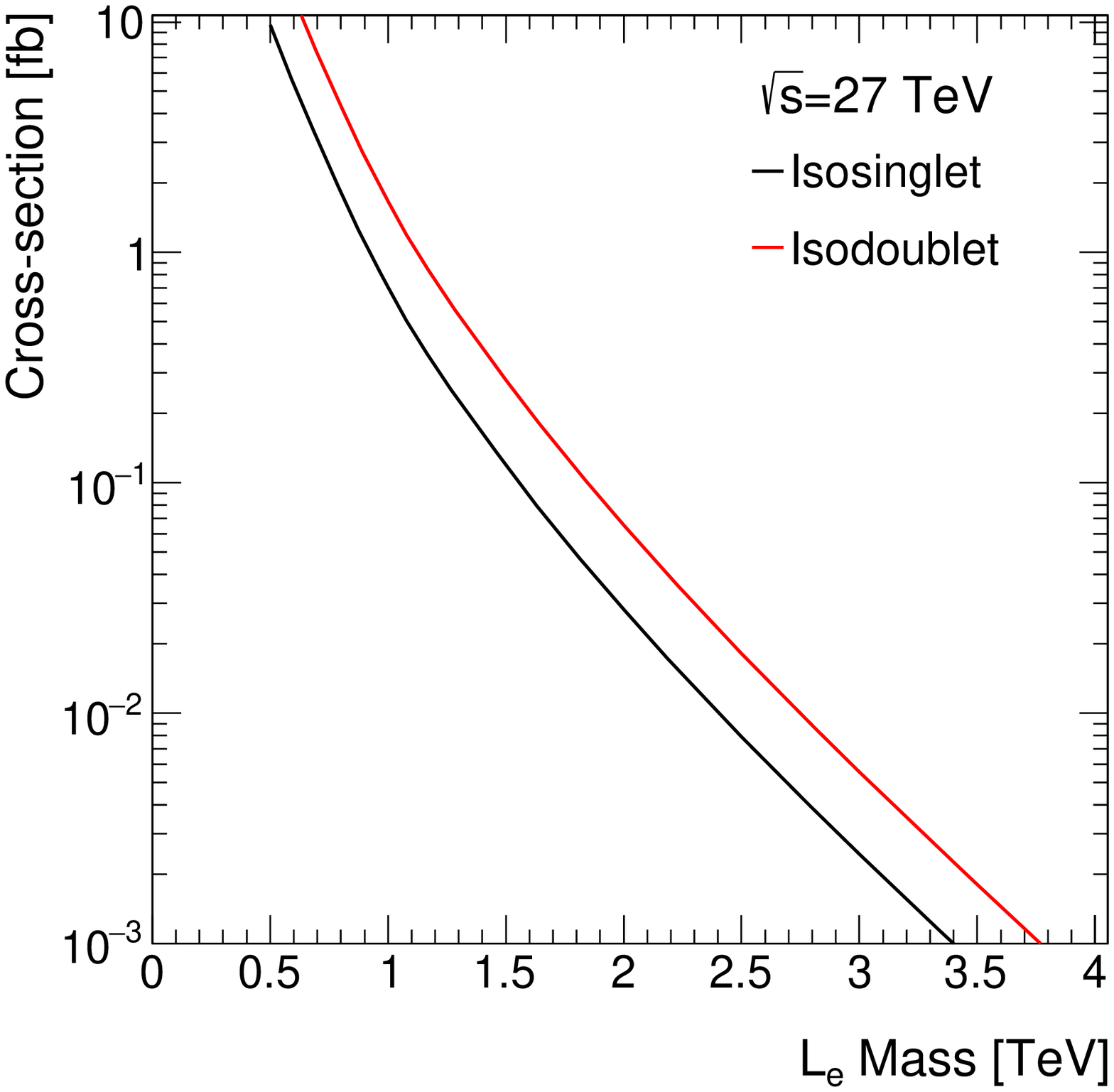}
  \end{minipage}
\caption{Cross-sections for pair production of charged vector-like leptons at the LHC and HE-LHC.}
\end{figure}
\begin{figure}[!h]
  \centering
    \includegraphics[width=0.49\textwidth]{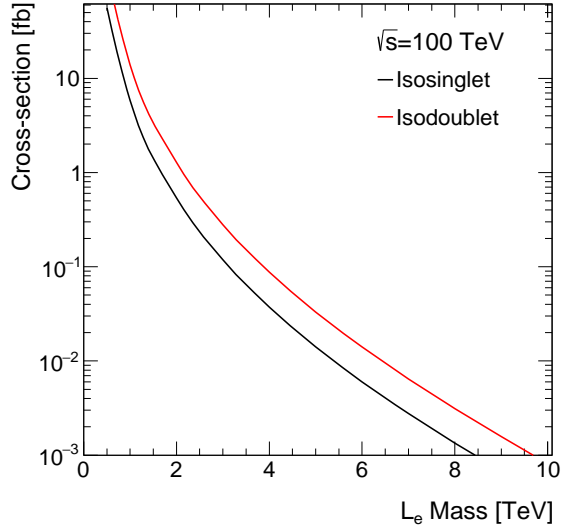}
\caption{Cross-sections for pair production of charged vector-like leptons at the FCC.}
\end{figure}
\newpage
	\section{Estimations of observation limits for specific channel}
	Generally speaking we have 3 decay channels for new lepton following by a lot of decay channels for W, Z and H bosons. In our opinion most promising signature for hadron colliders will be provided by decay of one of the leptons to Ze and another to He channel, followed by $H \rightarrow bb$ and $Z \rightarrow \mu^+ \mu^-$ decays. This decay chain gives opportunity to avoid complications coupled with missing energy, multi-lepton combinatorics and so on.
	\subsection{Hadron colliders}
	Cross-section for this channel is given by
	\begin{equation}
	\begin{split}
	\sigma(pp \rightarrow e^+e^-\mu^+\mu^-bb+X)=&2*Br(L \rightarrow He)*Br(L \rightarrow Ze)*(Br(H \rightarrow bb)* \\ &Br(Z \rightarrow \mu^+\mu^-)*\sigma(pp \rightarrow L^+_e L^-_e +X)
	\end{split}
	\end{equation}
According to the PDG Br($H \rightarrow bb$)=0.58 and Br($Z \rightarrow \mu^+\mu^-$)=$3.36*10^{-2}$, Br($L \rightarrow He$) and Br($L \rightarrow Ze$) are presented in Fig. 1, $\sigma(pp \rightarrow L^+_e L^-_e + X)$ is shown in Fig. 2 and Fig. 3. Assuming 25 events as observation limit for pair production at pp colliders we obtain achievable mass values at HL-LHC, HE-LHC and FCC presented in Table 2.
As mentioned above in the case of iso-doublet leptons $E_R-e_R$ mixing results in $E \rightarrow eZ/eH$ decays proportional to $sin^2\phi_R$, whereas $E_L-e_L$ mixing leads to $E \rightarrow W\nu_e$ decays proportional to $sin^2\phi_L$. In Fig. 4 (left) we present observable mass limits at HE-LHC for general case where vertical axis shows BR($E \rightarrow eZ$) = BR($E \rightarrow eH$) values. BR($E \rightarrow eZ$) = BR($E \rightarrow eH$) = 0.5 corresponds to $sin^2\phi_L$ = 0. Similar graphic for FCC is also shown in Fig. 4 (right).

	\begin{table}[ht]
		\centering
		\caption{Observable mass values for isosinglet and isodoublet (NC modes only) charged vector-like leptons.}
		\begin{tabular}{|c|cc|cc|cc|}
			\hline
							& \multicolumn{2}{c|}{HL-LHC}   & \multicolumn{2}{c|}{HE-LHC} &  \multicolumn{2}{c|}{FCC} \\
			$\mathcal{L} [fb^{-1}]$      & 1000 	& 3000 		& 1000 		&10000	&1000 	&20000	\\
			\hline
			Isosinglet M [GeV]            & 340 	& 460	 		& 490	 		&910		&845	 	&2025		\\
			\hline
			Isodoublet M [GeV]          & 610 	& 775	 		& 900	 		&1530		&1645	 	&3650		\\
			\hline
		\end{tabular}
		\label{tbl:1}
	\end{table}

\begin{figure}[!h]
  \centering
  \begin{minipage}[b]{0.49\textwidth}
    \includegraphics[width=\textwidth]{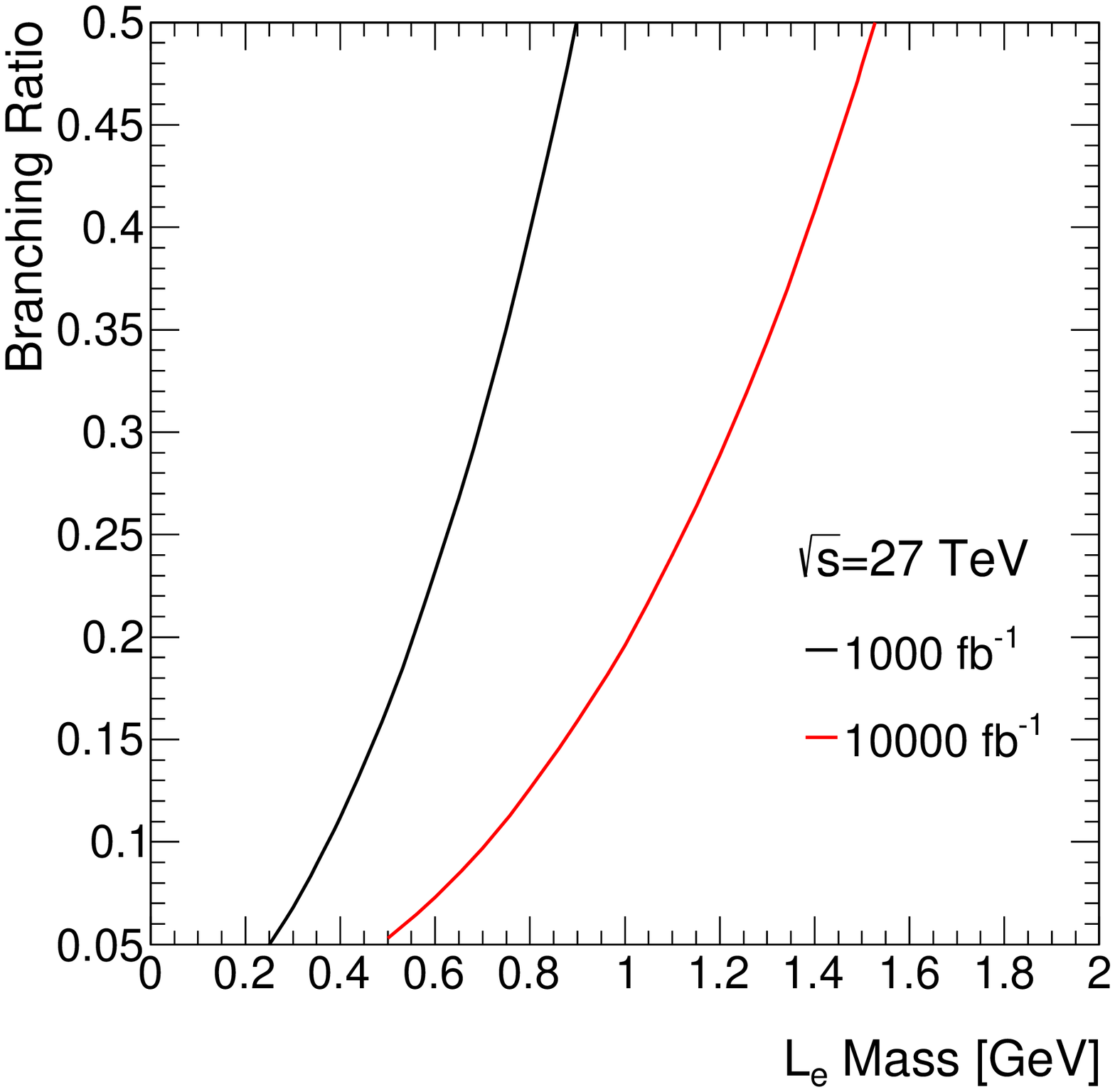}
  \end{minipage}
  \hfill
  \begin{minipage}[b]{0.49\textwidth}
    \includegraphics[width=\textwidth]{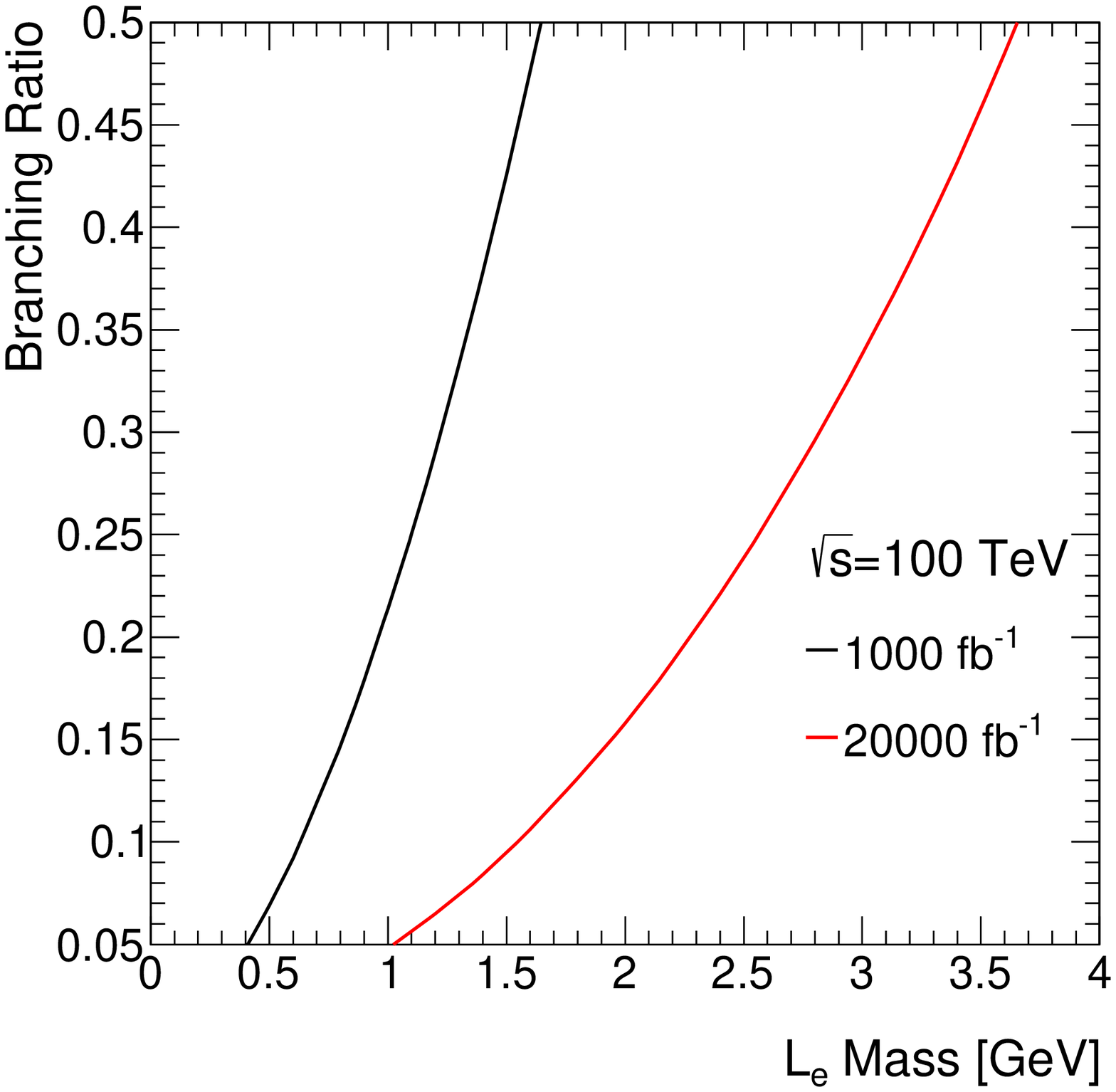}
  \end{minipage}
\caption{Observable mass limits for pair production of iso-doublet vector-like charged leptons at HE-LHC (left) and FCC (right) for general case, namely, $sin^2\phi_L \ne 0$.}
\end{figure}
Let us emphasize that these results are applicable for second generation charged VLL if $L_\mu \rightarrow \mu Z/\mu H$ modes are considered with subsequent $H \rightarrow bb$ and $Z \rightarrow e^+e^-$ decays.
	\subsection{Lepton colliders}
	At lepton colliders masses of iso-singlet and iso-doublet charged VLL can be scanned up to values close to kinematical limit, $\sqrt{s}/2$. Concerning signature under-consideration ILC with $\sqrt{s}$ = 1 TeV is comparable with the HL-LHC for isosinglet charged leptons, CLIC or Muon Collider (MC) with $\sqrt{s}$ = 3 TeV potential exceeds that of HL-LHC both for isosinglet and isodoublet and HE-LHC for isosinglet charged leptons, MC with $\sqrt{s}$ = 6 TeV potential exceeds that of HE-LHC both for isosinglet and isodoublet and FCC for isosinglet charged leptons.
	
	\section{Conclusion}
From phenomenology viewpoint vector-like leptons have the same status vector-like quarks and should be included into the new physics search programs of energy frontier colliders. Existence of iso-singlet quarks and vector-like charged leptons is favored by Flavor Democracy Hypothesis, which could provide an important key to solve the mystery of mass and mixing patterns of the SM fermions.
We argued that most promising signature for pair production of first family vector-like lepton at hadron colliders will be provided by decay of one of the leptons to $Ze$ and another to $He$ channel, followed by $H \rightarrow bb$ and $Z \rightarrow \mu^+ \mu^-$ decays. Summary of our results are presented in Table 2 and Figure 4. For example, HL-LHC will cover mass range up to 460 GeV for iso-singlet and 775 GeV for iso-doublet VL charged leptons. FCC could extend these ranges up to 2 TeV and 3.6 TeV, respectively.
Mass limits given above are results of rough estimations. Detailed simulations should be made for more realistic values. We assume that the discovery limits for pp colliders would be higher if other decay channels were also included in the analysis. Corresponding studies continue in this context.
	\newpage
	\bibliographystyle{unsrt}
	\bibliography{ms}

\begin{thebibliography}{10}

\bibitem{aad2012observation}
Georges Aad, Tatevik Abajyan, B~Abbott, J~Abdallah, S~Abdel Khalek, Ahmed~Ali
  Abdelalim, R~Aben, B~Abi, M~Abolins, OS~AbouZeid, et~al.
\newblock Observation of a new particle in the search for the standard model
  higgs boson with the {ATLAS} detector at the {LHC}.
\newblock {\em Physics Letters B}, 716(1):1--29, 2012.

\bibitem{chatrchyan2012observation}
Serguei Chatrchyan, Vardan Khachatryan, Albert~M Sirunyan, Armen Tumasyan,
  Wolfgang Adam, Ernest Aguilo, Thomas Bergauer, M~Dragicevic, J~Er{\"o},
  C~Fabjan, et~al.
\newblock Observation of a new boson at a mass of 125 {GeV} with the {CMS}
  experiment at the {LHC}.
\newblock {\em Physics Letters B}, 716(1):30--61, 2012.

\bibitem{sultansoy1998four}
S~Sultansoy.
\newblock Four ways to {TeV} scale.
\newblock {\em Turkish Journal of Physics}, 22(7):575--594, 1998.

\bibitem{ssultansoytalk1}
{S. Sultansoy}.
\newblock Four remarks on physics at {LHC}, invited talk at {ATLAS} week.
\newblock \textsc{url:}~\url{http://inspirehep.net/record/971257}, May 1997.

\bibitem{mchalm}
{M. Chalmers}.
\newblock Model physicist, {CERN} courier.
\newblock \textsc{url:}~\url{http://cerncourier.com/cws/article/cern/70138},
  2017.

\bibitem{de2017archeology}
A~De~R{\'u}jula.
\newblock Archeology and evolution of {QCD}.
\newblock In {\em EPJ Web of Conferences}, volume 137, page 01007. EDP
  Sciences, 2017.

\bibitem{sultansoy2007flavor}
Saleh Sultansoy.
\newblock Flavor democracy in particle physics.
\newblock In {\em AIP Conference Proceedings}, volume 899, pages 49--52.
  American Institute of Physics, 2007.

\bibitem{cetin2013qcd}
SA~Cetin, S~Sultansoy, and G~{\"U}nel.
\newblock Why {QCD} explorer stage of the {LHeC} should have high (est)
  priority.
\newblock {\em arXiv preprint arXiv:1305.5572}, 2013.

\bibitem{fernandez2012large}
JL~Abelleira Fernandez, C~Adolphsen, AN~Akay, H~Aksakal, JL~Albacete,
  S~Alekhin, P~Allport, V~Andreev, RB~Appleby, E~Arikan, et~al.
\newblock A large hadron electron collider at {CERN} report on the physics and
  design concepts for machine and detector.
\newblock {\em Journal of Physics G: Nuclear and Particle Physics},
  39(7):075001, 2012.

\bibitem{doi:10.1142/S0217751X10049165}
A.~N. Akay, H.~Karadeniz, and S.~Sultansoy.
\newblock Review of linac–ring-type collider proposals.
\newblock {\em International Journal of Modern Physics A}, 25(24):4589--4602,
  2010.

\bibitem{Acar:2016rde}
Y.~C. Acar, A.~N. Akay, S.~Beser, A.~C. Canbay, H.~Karadeniz, U.~Kaya, B.~B.
  Oner, and S.~Sultansoy.
\newblock {Future circular collider based lepton\textendash{}hadron and
  photon\textendash{}hadron colliders: Luminosity and physics}.
\newblock {\em Nucl. Instrum. Meth. A}, 871:47--53, 2017.

\bibitem{canbay2017sppc}
Ali~Can Canbay, Umit Kaya, Bora Ketenoglu, Bilgehan~Baris Oner, and Saleh
  Sultansoy.
\newblock Sppc based energy frontier lepton-proton colliders: luminosity and
  physics.
\newblock {\em Advances in High Energy Physics}, 2017, 2017.

\bibitem{sdufeffd468814}
Ümit Kaya, Bora Ketenoğlu, and Saleh Sultansoy.
\newblock The {LHeC} project: {e-Ring} revisited.
\newblock {\em Süleyman Demirel Üniversitesi Fen Edebiyat Fakültesi Fen
  Dergisi}, 13:173 -- 178, 2018.

\bibitem{kaya2019main}
U~Kaya, B~Ketenoglu, S~Sultansoy, and F~Zimmermann.
\newblock Main parameters of hl-lhc and he-lhc based mu-p colliders.
\newblock {\em arXiv preprint arXiv:1905.05564}, 2019.

\bibitem{greensite2003confinement}
Jeff Greensite.
\newblock The confinement problem in lattice gauge theory.
\newblock {\em Progress in Particle and Nuclear Physics}, 51(1):1--83, 2003.

\bibitem{arik2003turkish}
Engin Arik et~al.
\newblock Turkish comments on" future perspectives in {HEP}".
\newblock {\em arXiv preprint hep-ph/0302012}, 2003.

\bibitem{sultansoy2012fourth}
Saleh Sultansoy.
\newblock The fourth generation, linac-ring type colliders, preons and so on.
\newblock {\em arXiv preprint arXiv:1208.3127}, 2012.

\bibitem{sultansoy2019energy}
Saleh Sultansoy.
\newblock Energy frontier lepton-hadron colliders, vector-like quarks and
  leptons, preons and so on.
\newblock {\em arXiv preprint arXiv:1901.00309}, 2019.

\bibitem{HARARI1978459}
Haim Harari, Hervé Haut, and Jacques Weyers.
\newblock Quark masses and cabibbo angles.
\newblock {\em Physics Letters B}, 78(4):459--461, 1978.

\bibitem{FRITZSCH199292}
Harald Fritzsch.
\newblock Light neutrinos, nonuniversality of the leptonic weak interaction and
  a fourth massive generation.
\newblock {\em Physics Letters B}, 289(1):92--96, 1992.

\bibitem{datta1993flavour}
Amitava Datta.
\newblock Flavour democracy calls for the fourth generation.
\newblock {\em Pramana}, 40(6):L503--L509, 1993.

\bibitem{CELIKEL1995257}
A.~Çelikel, A.K. Çiftçi, and S.~Sultansoy.
\newblock {A search for the fourth SM family}.
\newblock {\em Physics Letters B}, 342(1):257--261, 1995.

\bibitem{kaya2018mass}
U~Kaya and S~Sultansoy.
\newblock {Mass Pattern of the SM Fermions: Flavor Democracy Revisited}.
\newblock {\em arXiv preprint arXiv:1801.03927}, 2018.

\bibitem{cline1}
{Cline D. and Soni A.}
\newblock Proceedings of the first international symposium on the fourthfamily
  of quarks and leptons.
\newblock Annals of the New York Academy of Sciences; 518., Feb. 1987.

\bibitem{cline2}
{Cline D. and Soni A.}
\newblock Proceedings of the second international symposium on the fourthfamily
  of quarks and leptons.
\newblock Annals of the New York Academy of Sciences; 578., Feb. 1989.

\bibitem{beyond1}
{Beyond the 3SM generation at the LHC era Workshop. [Document on homepage].;
  CERN, Geneva, Switzerland}.
\newblock \textsc{url:}~\url{http://indico.cern.ch/event/33285}, Sep. 2008.

\bibitem{beyond2}
{Second Workshop on Beyond 3 Generation Standard Model New Fermions at the
  Crossroads of Tevatron and LHC. [Document on homepage]. Taipei, Taiwan}.
\newblock \textsc{url:}~\url{http://indico.cern.ch/event/68036}, Jan. 2010.

\bibitem{beyond3}
{Third Workshop on Beyond 3 Generation Standard Model Under the light of the
  initial LHC results. [Document on homepage]. 2011 October 23-25; Istanbul,
  Turkey}.
\newblock \textsc{url:}~\url{https://indico.cern.ch/event/150154}, Oct. 2011.

\bibitem{holdom2009four}
Bob Holdom, WS~Hou, Tobias Hurth, Michelangelo~L Mangano, Saleh Sultansoy, and
  Gokhan {\"U}nel.
\newblock Four statements about the fourth generation.
\newblock {\em PMC Physics A}, 3(1):1--12, 2009.

\bibitem{cetin2011status}
SA~Cetin, GW-S Hou, VE~{\"O}zcan, AN~Rozanov, and S~Sultansoy.
\newblock Status of the fourth generation-a brief summary of b3sm-iii workshop
  in four parts.
\newblock {\em arXiv preprint arXiv:1112.2907}, 2011.

\bibitem{sultansoy2000four}
S~Sultansoy.
\newblock Why the four sm families.
\newblock {\em arXiv preprint hep-ph/0004271}, 2000.

\bibitem{PhysRevD.83.054022}
M.~Sahin, S.~Sultansoy, and S.~Turkoz.
\newblock Search for the fourth standard model family.
\newblock {\em Phys. Rev. D}, 83:054022, Mar 2011.

\bibitem{bar2013two}
Shaouly Bar-Shalom, Michael Geller, Soumitra Nandi, and Amarjit Soni.
\newblock Two higgs doublets, a 4th generation and a 125 gev higgs: a review.
\newblock {\em Advances in High Energy Physics}, 2013, 2013.

\bibitem{banerjee2014higgs}
Shankha Banerjee, Mariana Frank, and Santosh~Kumar Rai.
\newblock Higgs data confronts sequential fourth generation fermions in the
  higgs triplet model.
\newblock {\em Physical Review D}, 89(7):075005, 2014.

\bibitem{CHANOWITZ1979402}
M.S. Chanowitz, M.A. Furman, and I.~Hinchliffe.
\newblock Weak interactions of ultra heavy fermions (ii).
\newblock {\em Nuclear Physics B}, 153:402--430, 1979.

\bibitem{atlas2015search}
Collaboration ATLAS, Hans~Peter Beck, Alberto Cervelli, Antonio Ereditato,
  Sigve Haug, Sonja Kabana, Federico Meloni, Geoffrey Andr{\'e}~Adrien Mullier,
  Klaus-Peter Pretzl, Maria~Elena Stramaglia, et~al.
\newblock {Search for production of vector-like quark pairs and of four top
  quarks in the lepton-plus-jets final state in pp collisions at s= 8 TeV with
  the ATLAS detector}.
\newblock {\em Journal of High Energy Physics}, 8(8):105, 2015.

\bibitem{201882}
A.M. Sirunyan, A.~Tumasyan, W.~Adam, F.~Ambrogi, E.~Asilar, T.~Bergauer,
  J.~Brandstetter, E.~Brondolin, et~al.
\newblock {Search for pair production of vector-like quarks in the $bWb^-W$
  channel from proton–proton collisions at s=13TeV}.
\newblock {\em Physics Letters B}, 779:82--106, 2018.

\bibitem{salam1}
Salam G. and Weiler A.
\newblock {The Collider Reach}.
\newblock \textsc{url:}~\url{ http://collider-reach.web.cern.ch/ }.

\bibitem{beser2016possible}
S~Beser, U~Kaya, BB~Oner, and S~Sultansoy.
\newblock Possible discovery channel for fourth chiral family up-quark at the
  lhc.
\newblock {\em arXiv preprint arXiv:1607.07623}, 2016.

\bibitem{belfatto2021ckm}
Benedetta Belfatto and Zurab Berezhiani.
\newblock Are the ckm anomalies induced by vector-like quarks? limits from
  flavor changing and standard model precision tests.
\newblock {\em arXiv preprint arXiv:2103.05549}, 2021.

\bibitem{mehdiyev2007prospects}
R~Mehdiyev, S~Sultansoy, G~Unel, and M~Yilmaz.
\newblock Prospects to search for e 6 isosinglet quarks in atlas.
\newblock {\em The European Physical Journal C}, 49(2):613--622, 2007.

\bibitem{sultansoy2008e6}
Saleh Sultansoy and Gokhan Unel.
\newblock The e6 inspired isosinglet quark and the higgs boson.
\newblock {\em Physics Letters B}, 669(1):39--45, 2008.

\bibitem{mehdiyev2008down}
R~Mehdiyev, A~Siodmok, S~Sultansoy, and G~Unel.
\newblock Down type isosinglet quarks in atlas.
\newblock {\em The European Physical Journal C}, 54(3):507--516, 2008.

\bibitem{aguilar2009identifying}
Juan~Antonio Aguilar-Saavedra.
\newblock Identifying top partners at lhc.
\newblock {\em Journal of High Energy Physics}, 2009(11):030, 2009.

\bibitem{10.1093/ptep/ptaa104}
Particle~Data Group, P~A Zyla, R~M Barnett, J~Beringer, O~Dahl, D~A Dwyer, D~E
  Groom, C~J Lin, K~S Lugovsky, E~Pianori, D~J Robinson, C~G Wohl, W~M Yao,
  K~Agashe, G~Aielli, B~C Allanach, C~Amsler, et~al.
\newblock {Review of Particle Physics}.
\newblock {\em Progress of Theoretical and Experimental Physics}, 2020(8), 08
  2020.
\newblock 083C01.

\bibitem{heavysaleh}
S.~F. Sultanov.
\newblock Heavy leptons and quarks in models of electroweak interactions with
  extended gauge simmetry, PhD Thesis,IHEP(Protnivo, Moscow Region) 1985.

\bibitem{bhattiprolu2019prospects}
Prudhvi~N Bhattiprolu and Stephen~P Martin.
\newblock Prospects for vectorlike leptons at future proton-proton colliders.
\newblock {\em Physical Review D}, 100(1):015033, 2019.

\bibitem{del2008effects}
F~Del~Aguila, J~De~Blas, and M~Perez-Victoria.
\newblock Effects of new leptons in electroweak precision data.
\newblock {\em Physical Review D}, 78(1):013010, 2008.

\end{thebibliography}

\end{document}